# Topological Flat Band and Parity-Time Symmetry in a Honeycomb Lattice of Coupled Resonant Optical Waveguides

Xue-Yi Zhu[1], Samit Kumar Gupta[1], Xiao-Chen Sun[1], Cheng He[1,4], Gui-Xin Li[2], Jian-Hua Jiang[3*], Ming-Hui Lu[1,4*], Xiao-Ping Liu[1,4], and Yan-Feng Chen[1,4]

[1]Department of Materials Science and Engineering, College of Engineering and Applied Sciences and National Laboratory of Solid State Microstructures, Nanjing University, Nanjing 210093, China

[2]Department of Materials Science and Engineering, Southern University of Science and Technology, Shenzhen, 518055, China,

[3]College of Physics, Optoelectronics and Energy, Soochow University, Suzhou, 215006, China

[4]Collaborative Innovation Center of Advanced Microstructures, Nanjing University, Nanjing 210093, China

*Correspondence to Ming-Hui Lu (luminghui@nju.edu.cn), Jian-Hua Jiang (joejhjiang@hotmail.com)

## Abstract:

Two-dimensional (2D) coupled resonant optical waveguide (CROW), exhibiting topological edge states, provides an efficient platform for designing integrated topological photonic devices. In this paper, we propose an experimentally feasible design of 2D honeycomb CROW photonic structure. The characteristic optical system possesses two-fold and three-fold Dirac points at different positions in the Brillouin zone. The effective gauge fields implemented by the intrinsic pseudo-spin-orbit interaction open up topologically nontrivial bandgaps through the Dirac points. Spatial lattice geometries allow destructive wave interference, leading to a dispersionless, nearly-flat energy band in the vicinity of the three-fold Dirac point in the telecommunication frequency regime. This nontrivial nearly-flat band yields topologically protected edge states. The pertinent physical effects brought about due to non-Hermitian gain/loss medium into the honeycomb CROW device are discussed. The generalized gain-loss lattice with parity-time symmetry decouples the gain and the loss at opposite zigzag edges, leading to purely gain or loss edge channels. Meanwhile, the gain and loss effects on the armchair boundary cancel each other, giving rise to dissipationless edge states in non-Hermitian optical systems. These characteristics underpin the fundamental importance as well as the potential applications in various optical devices such as polarizers, optical couplers, beam splitters and slow light delay lines.

# Introduction:

The exploration of various topological states in quantum systems mainly focuses on condensed-matter and artificial photonic systems[1-3]. Optical topological states may have important applications in optical devices due to its characteristics of backscattering immunity. By breaking time-reversal symmetry using external magnetic fields, gapless topological edge states can emerge in gyromagnetic photonic crystals[4-6]. Alternatively, topological edge states can be realized by introducing effective magnetic gauge fields originating from periodic time-dependent[7] or spatial z-direction[8] modulations. However, the integration of time-dependent and z-direction modulations into semiconductor platforms is still quite challenging, while magnetic fields are incompatible with integrated miniature optical chips. Recently, 2D coupled resonant optical waveguide (CROW) which exhibits topologically protected edge state has been theoretically proposed[9, 10] and experimentally demonstrated[11, 12]. 2D dielectric CROWs, which can be fabricated using standard nanofabrication processes and are compatible with conventional photonic integrated circuits[11, 12], pave promising ways toward manipulation of light through topological edge states.

In this Letter, we design a 2D honeycomb CROW photonic structure which supports both two-fold and three-fold Dirac points at different high symmetry points in the Brillouin zone. The designed lattice geometry gives rise to destructive wave interference which results in a topologically nontrivial, nearly flat photonic band in the telecommunication frequency regime. The effective gauge fields implemented by the pseudo-spin-orbit interaction in the CROW opens up bandgaps with nontrivial topology through the Dirac points. The collective localization in the nearly-flat band depends on the lattice geometry design, impurities or structural disorders[13-18]. These modes have a large modal area, similar to the bulk propagating ones, but with a much lower group velocity. Gapless edge states between the nearly-flat band and the adjacent energy bands are revealed. The field distributions of CROW structure confirm the robustness of topological edge states. A 2D tight-binding model is then set up to analyze both the bulk and edge spectra, showing consistency with numerical simulations. On the other side, non-Hermitian photonic systems have drawn much research interest[19-23]. Such systems possess an anomalous symmetry-breaking transition[24], giving rise to abundant intriguing physical phenomena such as unidirectional transmission[25], asymmetric mode conversion[26] , PT laser[27, 28] and so on[29, 30]. The pertinent physical effects induced via introducing the gain/loss medium into the honeycomb CROW structure with PT symmetry are studied here. Remarkably, the armchair edges of the honeycomb CROW have balanced loss and gain, giving rise to dissipationless edge states. Such a phenomenon demonstrate the compatibility between topology and PT symmetry under certain conditions. In contrast, the zigzag edges break the gain and loss balance, leaving purely loss or gain edge states at the two opposite boundaries. Based on these intriguing properties, the proposed honeycomb CROW structure may play a crucial novel role in the optical devices such as slow light device, topological cavity, large-mode-area laser and so on. In the viewpoint of general wave theory, the underlying physics and exotic phenomena can be extended to other regimes, such as acoustics[31-34], plasmonics[35] and other systems.

## Manuscript:

The designed honeycomb CROW structure is shown in Fig. 1(a). The ring waveguides with refractive index $n = 3$ are periodically arranged in the *x-y* plane. The topological edge states appear at the boundary of the structure, as shown by the red curves in the figure. The part of the white dashed box in Fig. 1(a) shows the unit cell of the honeycomb lattice, while the detailed structure is shown in Fig. 1(b). In this structure, two identical optical ring waveguides occupy the "site" positions of the graphene-like configuration (denoted as "site" rings with an inner diameter $R_s = 1.653 um$). The optical ring waveguide supports both clockwise and counterclockwise modes, which constitute the two pseudo-spins [pseudo-spin up (down) denote the clockwise (anticlockwise) mode] in optical quantum spin Hall effect[10, 11, 36, 37]. The coupling between the clockwise and counterclockwise modes in an isolated ring is forbidden by time-reversal and rotation symmetries. Each site ring waveguide is connected with its nearest-neighbor site-ring waveguides through three coupling waveguides which are denoted as "link" rings with the same inner diameter $R_l = 1.653 um$. By designing directional coupling between two adjacent rings, the optical mode in the link ring always undergoes opposite rotation as compared to when it is in the site ring. As shown in Fig. 1(c), the widths (outer radius minus inner radius) of both the site and link rings are the same $w = 0.2 um$. A small space gap is assigned between adjacent rings $g = 0.1 um$, allowing efficient coupling between them. Using commercial FEM software (COMSOL MULTIPHYSICS), we obtain the transmission/reflection spectra by calculating the energy flux of the right output (T), as shown in Fig. 1 (d). In the frequency range of 190.0 THz to 196.0 THz, the transmission of the ring resonator is over 20%. Such efficient coupling gives rise to the photonic energy bands in the CROW lattice.

A tight-binding model is exploited to describe the honeycomb CROW lattice, which results in the following Hamiltonian[11, 12, 18]:

$$H_0 = \sum_{<ij>} t \cdot c_i^\dagger c_j \tag{1}$$

where $c_i^\dagger$ and $c_i$ are the photon creation and annihilation operators at lattice site $i$, respectively, $t$ is the tunnel coupling between two adjacent rings. The photonic band structure of the CROW system can be obtained from the Hamiltonian in wavevector space $\mathbf{H} = \sum_\mathbf{k} \Psi_\mathbf{k}^\dagger H_\mathbf{k} \Psi_\mathbf{k}$, where $\Psi_k = (c_{1\mathbf{k}}, c_{2\mathbf{k}}, c_{3\mathbf{k}}, c_{4\mathbf{k}}, c_{5\mathbf{k}})^T$ and

$$H_\mathbf{k} = \begin{pmatrix} 0 & e^{-i\mathbf{k}\cdot\mathbf{a}_2/2} & 0 & e^{i\mathbf{k}\cdot\mathbf{a}_2/2} & 0 \\ e^{i\mathbf{k}\cdot\mathbf{a}_2/2} & 0 & e^{i\mathbf{k}\cdot\mathbf{a}_1/2} & 0 & e^{i\mathbf{k}\cdot\mathbf{a}_3/2} \\ 0 & e^{-i\mathbf{k}\cdot\mathbf{a}_1/2} & 0 & e^{i\mathbf{k}\cdot\mathbf{a}_1/2} & 0 \\ e^{-i\mathbf{k}\cdot\mathbf{a}_2/2} & 0 & e^{-i\mathbf{k}\cdot\mathbf{a}_1/2} & 0 & e^{-i\mathbf{k}\cdot\mathbf{a}_3/2} \\ 0 & e^{-i\mathbf{k}\cdot\mathbf{a}_3/2} & 0 & e^{i\mathbf{k}\cdot\mathbf{a}_3/2} & 0 \end{pmatrix} \quad (2)$$

The three nearest neighbor vectors of the honeycomb structure (site rings) are given by $\mathbf{a}_1 = (0,1)$, $\mathbf{a}_2 = (\frac{\sqrt{3}}{2}, -\frac{1}{2})$, $\mathbf{a}_3 = (-\frac{\sqrt{3}}{2}, -\frac{1}{2})$. The energy bands of the Hamiltonian are obtained by diagonalizing the above matrix (for more details of the band structure, see the Supplemental Materials). At the middle of the photonic spectrum, a nearly-flat band appears across the entire Brillouin zone and the wave functions associated with the flat band have zero amplitude at link rings, demonstrating collective localization phenomenon. Indeed, as detailed in the Supplemental Materials, the Hamiltonian matrix can be block-diagonalized into two subsets: one subset constitutes only site-rings, while the other consists of only link-rings. An intrinsic pseudo-spin-orbit coupling term $H_{SO} = i\lambda_{so} \sum_{<ij>\alpha\beta} (\mathbf{d}_{ij}^1 \times \mathbf{d}_{ij}^2) \cdot \sigma_{\alpha\beta} c_{i\alpha}^\dagger c_{j\beta}$ is introduced to account for next-nearest-neighbor hopping and the quantum spin Hall effect in the honeycomb lattice[36, 37]. Here the $\lambda_{SO}$ corresponds to the next-nearest-neighbor (NNN) coupling strength[11], $\mathbf{d}_{ij}^1$ and $\mathbf{d}_{ij}^2$ are two vectors connecting the NNN site ring, and $\sigma$ stands for the Pauli matrices for the pseudo-spins. In the vicinity of the high symmetry points, the effective Hamiltonians become:

$$\begin{aligned} h_p^K &= v_K(\mathrm{p}_x \sigma_1 + \mathrm{p}_y \sigma_2) - \frac{\sqrt{3}}{2}\alpha\lambda_{so}\sigma_3 \\ h_p^\Gamma &= v_\Gamma(\mathrm{p}_x S_1 + \mathrm{p}_y S_2) - 2\sqrt{3}\lambda_{so} S_3 \end{aligned} \quad (3)$$

where $\mathbf{p} = \mathbf{k} - \mathbf{K}_\pm(\Gamma_o)$ and $\alpha$ is the spin index. This pseudo-spin-orbit coupling term opens up gaps in the bulk bands (Figure S1(b)) associated with the nontrivial $\mathbf{Z_2}$ index as shown in the Supplemental Materials. We simulate and obtain the photonic bands numerically which are shown in Figure 2(a). The existence of the flat band coincides with the picture illustrated in the tight-binding model.

The bulk and edge photonic spectrum of a strip of the honeycomb CROW structure is presented in Fig. 2(b) for zigzag edges. The gapless edge states in the band diagram are for the upper and lower boundaries for the pseudo-spin up mode [shown in Fig. 2(b) as the red and blue dots], respectively.

Through full-wave simulation, we obtain the field distribution of the topologically protected edge states in a finite-size lattice of 7×4 unit-cells. The excitation port is on the right-lower corner of the structure, as shown in the inset figure of Fig. 3(a). The photonic edge states are shown to be robust against both site and link defects [Figs. 3(b), (c) and (d)]. For the site defect, the boundary field distribution is shown in Figure 3(b), where the inset represents the specific location of the site defect. Robust unidirectional light propagation around the site defect is observed. Similarly as shown in Figure 3(c), we observe unidirectional light propagation around the defect. Besides, the edge state can also be robust against an additional site ring [Figure 3(d)].

In non-Hermitian photonic topological insulators a potent question has been whether it can support topological quantum edge states, while its Hermitian analogues exhibit the same based on quantum Hall effects and topological insulators. Subsequently, a number of studies[38-41] explored and confirmed this possibility. In the hexagonal CROW structure, we now consider the introduction of gain and loss medium to render it to its non-Hermitian configuration. In our design, one of the site rings of honeycomb CROW has a refractive index of $3+0.005i$ (gain), while the other has a refractive index of $3-0.005i$ (loss). All link rings have no gain or loss [see Fig. 4(a)]. As shown in Fig. 4(b), the real part of the bulk and edge spectrum remains the same for zigzag edges. The gain and loss are generally not balanced. Interestingly, for zigzag edges, the gain and loss are split for the two opposite boundaries: one boundary the edge states undergo gain, on the other boundary the edge states undergo loss. As shown in Fig. 4(e), the imaginary parts of the eigenvalues are split up into two, one of which is positive corresponding to the loss boundary, while the other negative, corresponding to the gain boundary [as also depicted in Fig. 4(a)] Remarkably, the armchair edge has balanced loss and gain, leading to lossless edge states. This observation demonstrate the compatibility between topological edge states and the parity-time symmetry, at least in certain conditions. The non-Hermitian response of different boundaries can be detected by transmission along these boundaries, as shown in Figs. 4(c) and 4(d). Fig. 4(c) shows that the upper boundary of zigzag edge states becomes pure gain (with transmission above the original transmission of the Hermitian system), while the low boundary becomes pure loss (with decreased transmission). In contrast, as shown in Fig. 4(d), the transmission along the armchair edge is identical for the non-Hermitian system as that for the original Hermitian system.

## Conclusions:

In conclusion, we have designed a honeycomb CROW structure containing a nontrivial nearly-flat band and topologically protected edge states. The Hermitian system has been modeled by a 2D tight binding framework which efficiently corroborates the energy band gap and the nontrivial nearly-flat band. In both the Hermitian and non-Hermitian honeycomb CROW structures the excitations of topologically protected edge states and pertinent effects have been elucidated. The non-Hermitian aspects of the photonic topological insulator systems may provide more controllable design freedom in achieving nontrivial physical effects. The topologically protected edge states can be used in the asymmetric selective control of light intensity in different boundaries. The collective localized modes formed due to the nearly-flat band possess large mode area and lower group velocity which may be applied to large mode area laser and slow light devices. These characteristics may be suggestive of potential application prospects and research values in optical devices such as optical couplers, beam splitters and so on. Additionally, the photonic systems in analogy to the quantum or electronic systems can be used for controlling some exotic optical properties, which can envisage an excellent optical platform for quantum simulations.

## Acknowledgments:

This work was supported by the National Key R&D Program of China (2017YFA0303702 and 2017YFA0305100), and the National Natural Science Foundation of China (Grant No.11625418, No. 11474158, No. 51732006, No. 51721001, No. 51472114 and No. 11675116). We also acknowledge the support of the Natural Science Foundation of Jiangsu Province (BK20140019)

**Figure captions:**

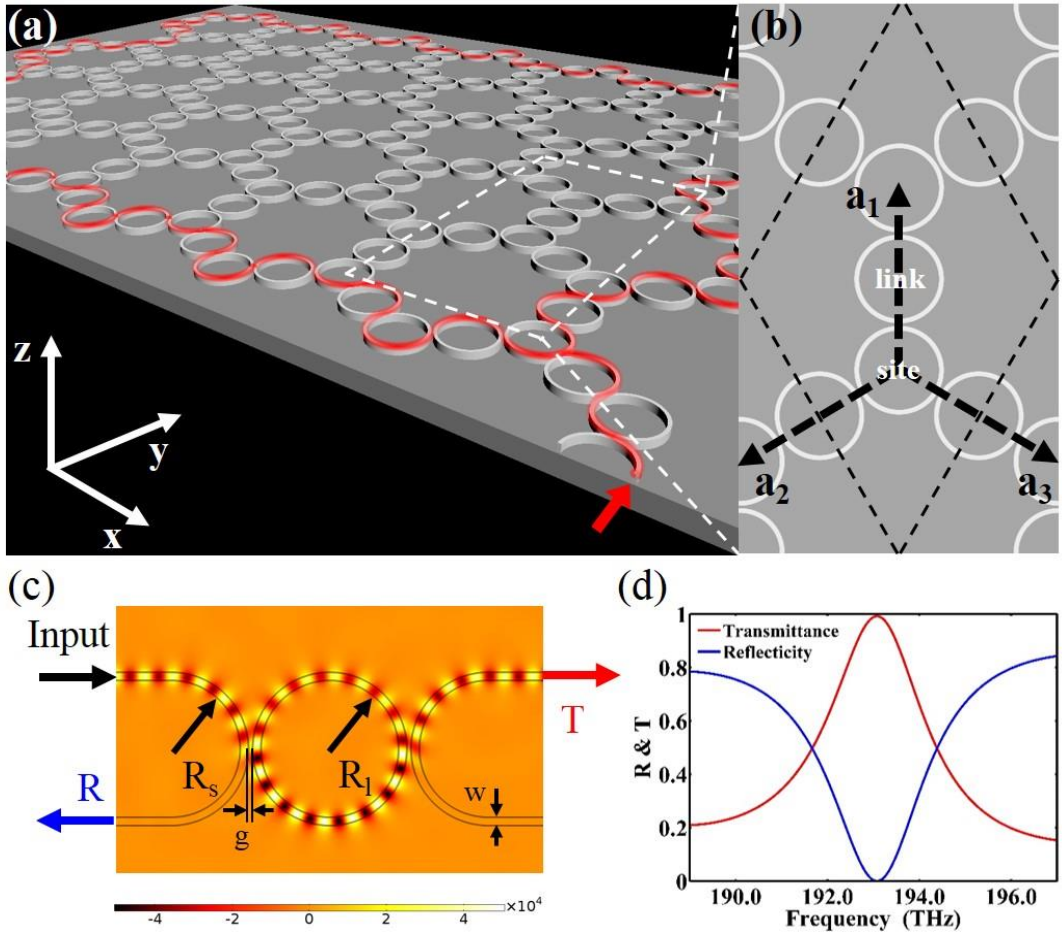

Fig. 1. (a) a schematic diagram of honeycomb CROW structures. The topological edge state are shown as the red line. (b) The detailed unit cell diagram of honeycomb CROW structures. The black dashed line represents a single repeating unit, and the arrows denote three nearest neighbor vectors $\mathbf{a_1} = (0,1)$, $\mathbf{a_2} = (\frac{\sqrt{3}}{2}, -\frac{1}{2})$, $\mathbf{a_3} = (-\frac{\sqrt{3}}{2}, -\frac{1}{2})$. (c) Configuration of site ring and link ring, with the same width $w = 0.2um$ and inner radii $R = 1.653um$. The separation is $g = 0.1um$. The incident, reflected and transmitted ports are represented by INPUT, R, T, respectively. (d) Transmission/Reflection spectra from 191.0 THz to 197.0 THz.

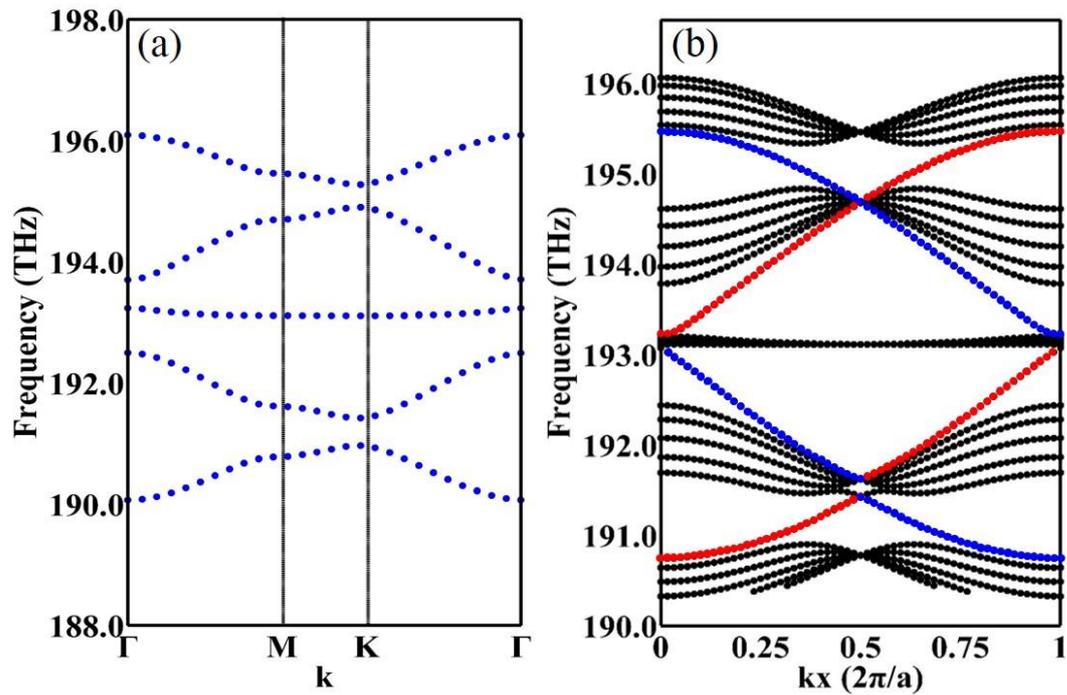

Fig. 2. (a) Schematic diagram of the energy band. (b) Schematic diagram of the projected band. The gapless edge states are shown in the red and blue dots in the figure, which corresponds to the upper and lower borders, respectively. The calculation interval of $k_x$ is $-\Gamma \to K \to \Gamma$ ($\Gamma$ and $K$ are high symmetric points in the first Brillouin zone), and $k_y$ is always zero in the calculation

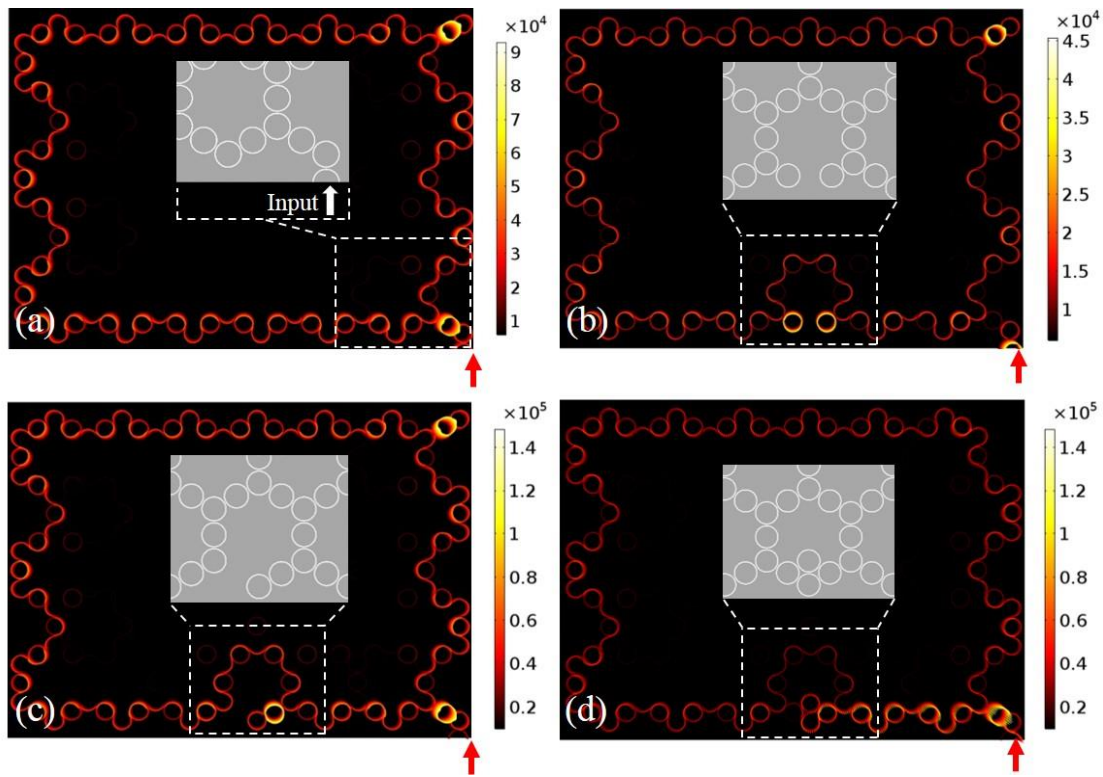

Fig. 3. The field distributions of topologically protected edge state and the bulk localized mode. (a) The field distribution of topologically protected edge state, white dashed box represents the input port of CROW structure.(b) (c) Robust edge state propagation with the site ring defect and the link ring defect, respectively. (d) Robust edge state propagation with an additional site ring.

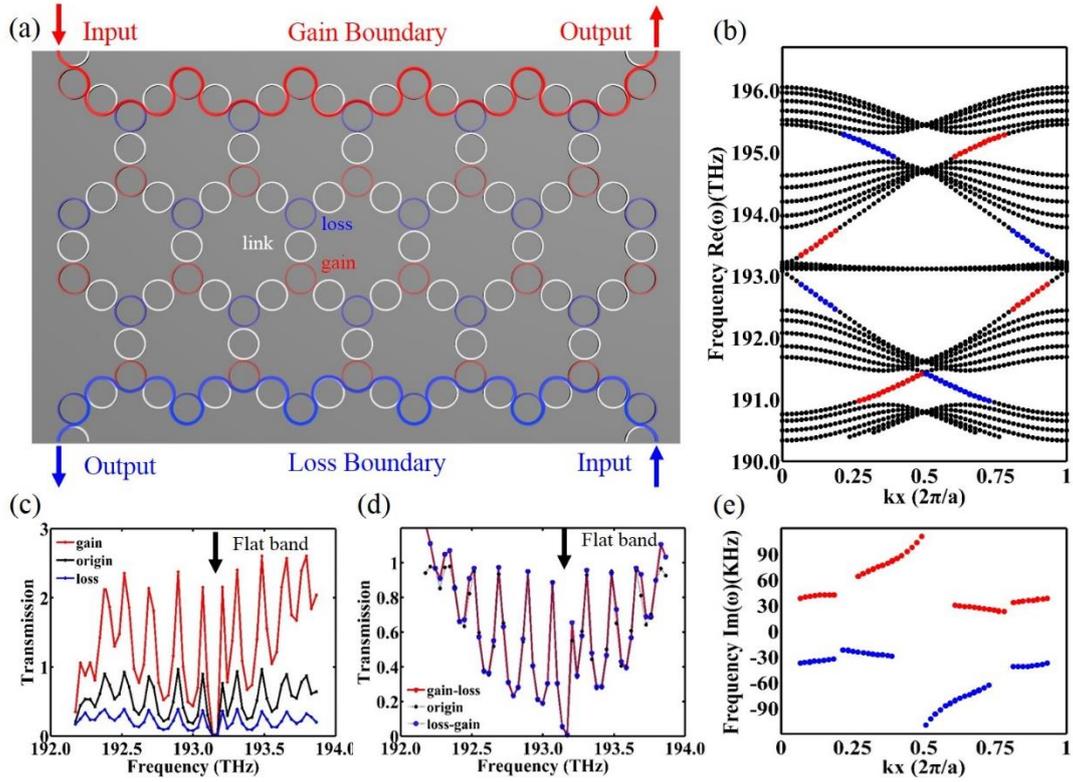

Fig. 4. (a) Schematic diagram of honeycomb CROW structure include gain/loss medium. The blue, red and white rings represent the loss medium ($n = 3+0.005i$), gain medium ($n = 3-0.005i$) and dielectric medium ($n = 3$), respectively. The blue and red line denote the loss and gain boundaries. And the arrows represent the energy input or output ports. (b)(e) The real and imaginary part of the projected band. The gapless edge states are shown in the red and blue dots in the figure, which corresponds the upper gain border and lower loss border, respectively. (c) The gain/loss energy response of the upper and lower zigzag boundaries. (d) The no gain and no loss energy response of the left and right armchair boundaries